\documentclass[final]{rmaa}
\usepackage{graphics}

\title{Romano's star in M\,33 - LBV candidate or LBV?}

\author{R.~Kurtev\altaffilmark{1}, O.~Sholukhova\altaffilmark{2}, J.~Borissova\altaffilmark{3} \and L.Georgiev\altaffilmark{4}}

\altaffiltext{1}{Department of Astronomy, Sofia University, Sofia, Bulgaria}
\altaffiltext{2}{Special Astrophysical Observatory, Russia}
\altaffiltext{3}{Institute of Astronomy, Bulgarian Academy of Sciences, Sofia, Bulgaria}
\altaffiltext{4}{Instituto de Astronom\'{\i}a, Universidad Nacional Aut\'onoma de M\'exico, M\'exico}

\fulladdresses{
\item Radostin Kurtev: Department of Astronomy, Faculty of Physics, Sofia University, 5 James Bourchier Blvd., BG\,--\,1164 Sofia, Bulgaria (kurtev@phys.uni-sofia.bg)
\item Olga Sholukhova: Special Astrophysical Observatory, Nizhnij Arkhyz, Karachai-Circessia 357147, Russia
(olga@sao.ru)
\item Jordanka Borissova: Institute of Astronomy, Bulgarian Academy of Sciences,
72~Tsarigradsko chauss\`ee, BG\,--\,1784 Sofia, Bulgaria (jura@haemimont.bg)
\item Leonid Georgiev: Instituto de Astronom\'{\i}a, Universidad Nacional Aut\'onoma de M\'exico, Apartado postal 70-264, 04510 M\'exico, D.F., M\'exico (georgiev@astroscu.unam.mx)
}

\shortauthor{Kurtev et al.}
\shorttitle{Romano's star in M\,33 - LBV candidate or LBV?}

\SetVolume{00} \SetFirstPage{1} \SetYear{2001}
\ReceivedDate{2000 May 26}
\AcceptedDate{2001 May 26}

\resumen{ Presentamos la curva de luz de la estrella GR\,290 (la estrella de
Romano) candidata de LBV en la galaxia M\,33. La fotometr\'{i}a se hizo
con  22 placas fotogr\'{a}ficas en banda $B$ de la galaxia M\,33 obtenidas en
el periodo 1982 -- 1990. Presentamos tambien CCD $B$ fotometr\'{i}a. El
an\'{a}lisis de nuestros datos junto con las magnitutes publicadas por
Romano (1978) muestran erupciones  "normales" con amplituda de $1$ magnidude
con escala de tiempo de $20$ a\~nos. Tambien se observa una variabilidad con
amplituda de $0.5$ magnitudes y per\'{i}odo de $320$ dias  aproximadamente
cual es el comportamiento tipico de los LBVs. }

\abstract{We present the light curve of Luminous  Blue Variable candidate star GR\,290
(Romano's star) in M\,33. The photographic photometry was made in photographic
plates taken in $B$ band of the M\,33 galaxy and cover an eight year period,
1982 -- 1990. Twenty five plates, separated in seven groups, have been used.
CCD $B$ magnitude of the star is also presented.  The analysis of our data together with
the Romano's magnitudes (1978) shows "normal"
eruptions with amplitude of more than $1$ mag and timescale of about $20$ years
and smaller oscillations with amplitude $0.5$ mag and a period of about 320
days. This is a typical photometrical behavior for LBVs.}

\keywords{stars: Luminous Blue Variables -- photometric history --
galaxies: individual: M33 -- galaxies: Local Group -- galaxies: stellar
content}

\begin{document}

\maketitle

\section{Introduction}
 Luminous Blue Variables (LBVs) form a group of  irregular variables
characterized by their high intrinsic luminosities.  LBVs are very short-lived
phase of massive star evolution between core hydrogen burning O-type stars and
helium-burning WR stars. LBV is a term coined by Conti (1984) that covers the
S\,Dor variables,
 the Hubble-Sandage variables and the P\,Cygni variable stars. The distinctive
characteristics of LBVs are the outbursts,
circumstellar nebulae and typical spectral features. There are only 33
"confirmed" and approx. 35 "candidate" LBVs located in 10 galaxies. Reviews of
LBVs can be found in Humphreys \& Davidson (1994) and in Nota \& Lamers, eds.
(1997).  The variabilities shown by the LBVs are of different amplitudes and
time scales. There are giant eruptions with amplitudes greater than 2 mag
but they are very rare, normal outbursts have amplitudes of 1 or 2 mag in
the optical bands and time scales of years or tens of years.  There are also
smaller quasi-periodic oscillations (cyclicities) with amplitudes of about
half a magnitude and microvariations of less than $0.1$mag.

The original paper of Hubble \& Sandage (1953) included four stars in
M\,33 (Vars. A, B, C and 2). All of them were distinguished by "blue" color
and irregular variability. Later, van den Bergh et al. (1975) added to this
list Var\,83. Romano (1978) discovered a variable star of Hubble-Sandage type
close to the external spiral arm of M\,33 designated as GR\,290 ("Romano's
star" is more popular). This star is classified as LBV-candidate in Humphreys
\& Davidson (1994) due only to reason of variability.

\begin{table*}[t]
\begin{center}
 \caption{Photometry in $B$ of Romano's star in M\,33}
\tabcolsep=2.3pt
\begin {tabular}{l l l @{\hspace{15pt}}l l l @{\hspace{15pt}} l l l
@{\hspace{15pt}} l l l}
\hline
JD    & $B$   &  $\sigma(B)$ & JD    &  $B$   &  $\sigma(B)$ & JD
&  $B$ &  $\sigma(B)$ & JD    &  $B$ &  $\sigma(B)$    \\
$244+$ & & & $244+$ & & & $244+$  & & & $244+$  & & \\
\hline
45286 & 17.16 & 0.15 & 45591 & 17.40 & 0.16 & 46707 & 17.46 & 0.16 & 48177 &
16.88 & 0.18 \\
45295 & 17.25 & 0.15 & 45623 & 17.51 & 0.23 & 46707 & 17.51 &
0.16 & 48180 & 16.96 & 0.17 \\
45296 & 17.19 & 0.23 & 45625 & 17.61 & 0.16 &
46708 & 17.30 & 0.07 & 48180 & 16.99 & 0.13 \\
45297 & 17.24 & 0.25 & 45702 &
17.11 & 0.08 & 46708 & 17.52 & 0.10 & 48180 & 16.89 & 0.12 \\
45588 & 17.35 &
0.23 & 45929 & 17.30 & 0.08 & 46709 & 17.49 & 0.08 & 51341$^\star$ & 17.35 &
0.03 \\
45588 & 17.34 & 0.26 & 45968 & 17.38 & 0.09 & 46738 & 17.45 & 0.11 & & & \\
45590 & 17.55 & 0.20 & 46435 & 17.55 & 0.08 & 46738 & 17.32 & 0.17 & & & \\
\hline
\end{tabular}
\label{Tab1}
\end{center}
\vspace{-7pt}
\hspace{2cm}$^\star$ -- {\small CCD $B$ magnitude (SAO 0.6-m telescope)}
\end{table*}

As an additional step in the process of confirmation or rejection of its LBV status,
we follow the  observational sequence of Romano (1978)
and to see whether the star has shown any later outbursts and
whether there is a periodicity in the light changes.

This photometric search is based on plates of M\,33 from the collection
of the Bulgarian National Astronomical Observatory (BNAO).

 The observational material and the photometric measurement techniques
are presented in section 2. In section 3 we present the light curve of
Romano's star, photometric behavior of residuals after removing the basic
trend of magnitude, and a short discussion.

\section{Observations and reductions}

The analysis of light changes of Romano's star is based on Romano's
observations, photographic observations with the 2 m Rozhen telescope
and CCD observation with the 0.6-m telescope of SAO (Russia).

 \subsection{Photographic observations}

A sample of photographic plates of M\,33 from the collection of the Bulgarian
National Astronomical Observatory -- Rozhen was used.  All plates
have been taken with the 2 m RCC f/8 Rozhen reflector.

We used twenty five 30 $\times$ 30 cm $B$-plates,  (103aO, IIaO and
ORWO ZU\,21 emulsions, GG\,385 glass filter).  The plates were taken
from November, 1982 to October, 1990. Julian days of observations and
$B$ magnitudes of Romano's star are
presented in Table\,1.  The plate scale is 12.8 arcsec mm$^{-1}$ and
the area covered is 1$^\circ$ $\times$ 1$^\circ$.  The whole image of
M\,33 fits in each plate.

The measurements have been made with a MF-4 densitometer with a constant
diaphragm at the Astronomical Observatory of the  Sofia University. At least
four estimations of sky background for each star were obtained and then an
averaged value was used. The calibration curves have been constructed using the
photoelectric sequence of  Sandage \& Johnson (1974).  A variety of functions
have been used to obtain the best fit of the data.  For each plate standard
deviations of measurements are presented in Table\,1.

\begin{figure}
\begin{center}
\includegraphics[width=\columnwidth]{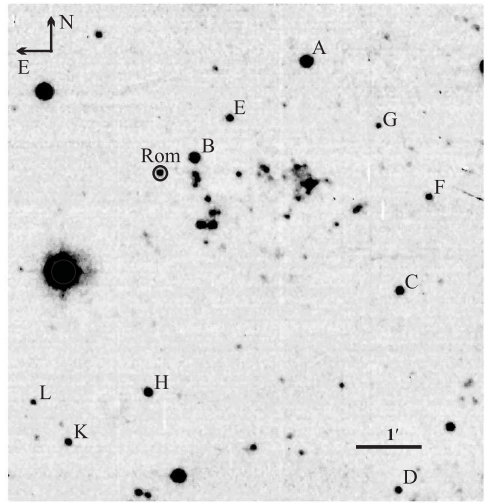}
\caption{ CCD $B$ image of the area around Romano's star.  North is
up and East is to the left. Reference stars A -- L from second photographic
sequence are shown.
  \label{Fig1}}
\end{center}
\end{figure}

\subsection{CCD observations}

In Table\,1 we present also CCD $B$ magnitude of Romano's star. The data were obtained on
June 21, 1999 with 525 $\times$ 600 CCD camera on the 0.6-m Zeiss telescope (Vlasiuk 1997) of
the Special Astrophysical Observatory -- SAO (Russia).  A standard $B$ filter was used.
The seeing during the observations was 2 -- 3 arcsec. The scale was 0.84 arcsec/pixel, resulting in
a field size of about 8 arcmin.  The photometry of the program frame was
carried out by PSF-fitting using { IRAF/DAOPHOT}. Transformation to the
standard system is based on average photographic $B$ magnitudes of reference
stars (A -- L in Fig.\,1) from the five best Rozhen $B$ plates. The average
magnitudes and root mean squares (r.m.s.) for these stars are given in
Table\,2.

\begin{table}[h]
\begin{center}
 \caption{Average $\langle B\rangle$ magnitudes and R.M.S. of
reference stars}
 \begin {tabular}{ l l l l l l }
 \hline
St & \multicolumn{1}{c}{$\langle B\rangle$} & R.M.S.(B) & St &
\multicolumn{1}{c}{$\langle B\rangle$} & R.M.S.(B) \\
 \hline
 A & 15.49 &
0.18 &    F & 17.33 & 0.06 \\
 B & 15.79 & 0.07 &   G & 17.86 & 0.03 \\
 C &
16.48 & 0.13 &    H & 16.33 & 0.16 \\
 D & 16.35 & 0.10 &     K & 17.10 & 0.19
\\
 E & 17.22 & 0.13 &    L & 17.91 & 0.07 \\
 \hline
 \end{tabular}
\end{center}
 \end{table}

\subsection{Reductions of the original Romano's data to the Johnson $B$
system}

The magnitudes of the variable star by Romano (1978) are obtained
by visual comparison with the stars from the Hubble \& Sandage
(1953) sequence and are in the old $\rm m_{ph}$ system. In order to compare
them with our data, Romano's observations were transformed to the Johnson $B$
system. This transformation was based on the twelve common sequence stars of
Hubble \& Sandage (1953) and Sandage \& Johnson (1974) -- 15, 16, 19, A2, A4,
A7, A10, A11, A12, A14, A16, and A17. The least squares fit gives:

 $$B=1.064\,{\rm m_{ph}}-0.831.$$
 with the fit standard error $\sigma=0.09$.

\begin{figure}
\begin{center}
\includegraphics[width=\columnwidth]{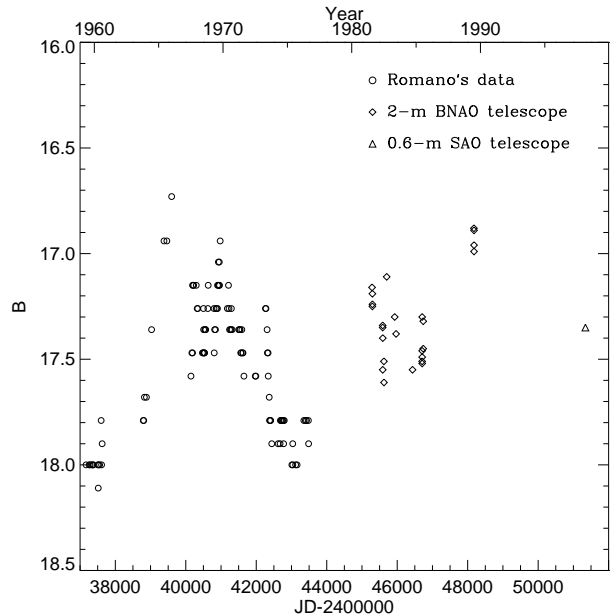}
\end{center}
\caption{The long term $B$ light curve of the Romano's star. Open circles
represent original Romano's observations, open diamonds -- 2-m Rozhen
observations, and open triangle -- 0.6-m SAO observation.
  \label{Fig2}}
\end{figure}

\section{Periodogram analysis and results}

The results of our photometry are given in Table~\ref{Tab1}. The light curve
of Romano's star is presented in Fig.\,2. The observations of Romano (1978)
transferred in $B$  are given by open circles, photographic $B$-magnitudes
from Rozhen 2-m RCC telescope -- by open diamonds, and the CCD $B$-magnitude
from 0.6-m  telescope -- an open triangle. It is seen that Romano's star
presents two maxima in the last $40$ years. The first one is around 1970 and
the second one is around 1990. There is a local increase of brightness
($B=16.7$) around 1967, formed from three separate observations. It does
not change the common appearance of the light curve. The modified {CLEAN}
algorithm for deconvolution of "dirty" Fourier spectrum of unequally spaced
data (Roberts et al. 1987) was applied in the search for multi-periodic
variability of the star. The derived mean period for the whole data set
(Romano's $+$ ours) by three methods is $6250\pm 30$ days.

\begin{figure}
\begin{center}
\includegraphics[width=\columnwidth]{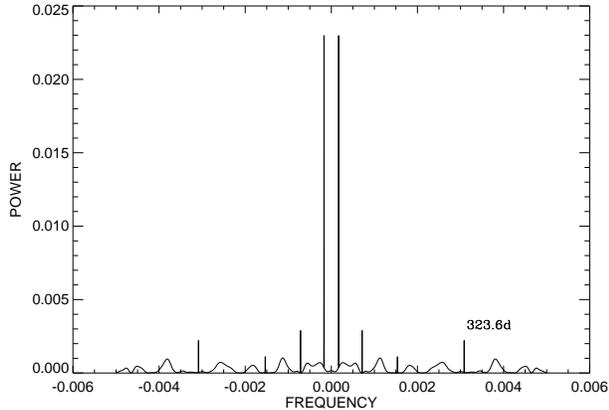}
\end{center}
\caption{CLEAN power spectrum of Romano's data (see Roberts et al.
1987 for details). The first peak at $f=1.6 10^{-4}{\rm d^{-1}}$ with amplitude $0.023$
corresponds to the mean period of $6250$ days. The second and third features (peaks at
$f=7.1 10^{-4}{\rm d^{-1}}$ and $f=1.55 10^{-3}{\rm d^{-1}}$)
correspond to the $1408$ and $645.2$ days pseudo-periodicities.  The fourth peak
represents a 323.6 days secondary period.
  \label{Fig3}}
\end{figure}

 Romano characterized the photometric behavior of the star as "irregular
variations between $16.5$ and $17.8$ pg". Looking on the light curve of
Romano (1978) however (open circles in Fig.\,2), one can find hints of
oscillations in brightness with smaller amplitude. The {CLEAN} algorithm
allows detection of multiple periodicity in the data set. Along with the mean
period there is a clear presence of  another much shorter period of $323.6\pm
0.1$ days. CLEAN'ed power spectrum of Romano's data (see Roberts et al. 1987 for
details) is given in Fig.\,3. The pseudo-periods of
$1408\pm 20$ ($f=7.1 10^{-4}{\rm d^{-1}}, {\rm S} =0.004$) and $645.2\pm 4.2$ days
($f=1.55 10^{-3}{\rm d^{-1}}, {\rm S} =0.001$) in Fig.\,3 do not lead to a reasonable light
curve. The basic trend of Romano's data (open circles) was fitted with cubic
spline. Removing the magnitude trend  we found the secondary periodicity of
the residuals, using a least-squares periodogram analysis by means of the
phase dispersion minimization (PDM) task available in {IRAF}, as well as a
period-finding program based on Lafler-Kinman's (1965) "theta" statistics
(LK). The obtained periods are $\rm 322.1\pm 0.2$ and $\rm 323.2\pm 0.1$ days
respectively and the average amplitude is about $0.4$ mag. The mean light
curve of the residuals is given in Fig.\,4 (upper panel).

The period used
for this mean curve is $323.6$ days -- obtained by CLEAN algorithm. The
amplitude is comparable with the scatter of the data ($0.3$ mag -- typical
for photographic photometry) but the presence of the period is obvious.

 It is difficult to fit and remove the basic trend of our subset of the data, because
of their relatively small number and unequal spacing. Despite of this,
attempt to find secondary periodicity of the residuals was made. The obtained
period from {CLEAN} is $263.16\pm 0.28$ days. LK gives the period of
$270.1$days. The mean light curve of the residuals obtained with the last
period is presented in Fig.\,4 (lower panel). These periods are more or less
speculative. More precise and much more equally distributed observations are
needed for reliable analysis, but in any case, in our subset of the data
there is presence of periodicity too.

Light curve of Romano's star is typical for LBVs and shows "normal"
eruptions with amplitude of more than $1$ mag and
timescale of about $20$ years, and smaller periodic oscillations. The presence
of smaller oscillations of about half a magnitude and timescales of months and
years on top of the  longer-term "normal" eruptions is one of the
"trade-marks" of many LBVs. Probably the closest case is the star AG Carinae
(Sterken et al. 1996).

\begin{figure}[h]
\begin{center}
\includegraphics[width=\columnwidth]{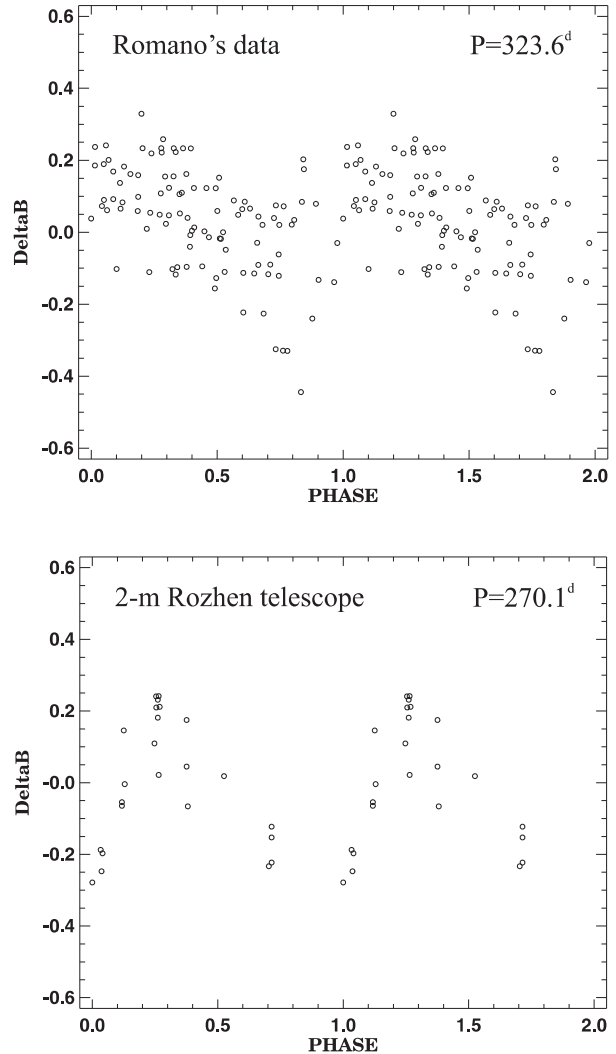}
\end{center}
\caption{The mean light curve of the residuals, obtained after removing
the basic trend in the magnitude of Romano's observations (upper panel). The
mean light curve of the residuals of our subset of the data (lower panel).  \label{Fig4}}
\end{figure}

 The photometric behavior of Romano's star in the last 40 years suggests that
the star should be considered an LBV. Additional spectroscopic investigation
and detection (or not) of a circumstellar nebula are necessary to finally
confirm its status.

\acknowledgements

 It is a pleasure to thank Prof. G. Ivanov, Dr. P. Kunchev and Dr. Ts. Georgiev
 for letting us use their plates of M\,33. This research was supported in part
 by the Bulgarian National Science Foundation grant under contract No. F-812/1998
 with the Bulgarian Ministry of Education and Sciences and CONACyT contract E130.1243/2000.
 O. Sholukhova is thankful for support RFBR grant No 00-02-16588.

\end{document}